# Trading zones revisited

## Collins, Harry[a], Evans, Robert[a], and Gorman, Michael[b]

[a] Cardiff University
[b] University of Virginia

**Abstract**

We describe and then elaborate the model of trading zones first presented in Collins et al 2007 – 'Trading Zones and Interactional Expertise'. We believe this expanded version of the model includes some very important but previously overlooked ways for separate language communities to communicate. We illustrate the argument with examples.

## 1 Introduction

In Collins, Evans and Gorman's 2007 paper – 'Trading Zones and Interactional Expertise' – interactions between separate linguistic communities, often known as 'trading zones', were shown to work in variety of different ways. The term 'trading zone' was introduced by Peter Galison as a supposed resolution of the problem caused by Kuhn's notion of 'paradigm incommensurability'. Under paradigm incommensurability, the concepts belonging to one paradigm cannot be translated into those of another paradigm. We tend to agree with Kuhn's characterisation of the problem but apply it far more generally to 'forms of life' which vary hugely in scale and are embedded within one another.[1] We refer to all cases in which there is tension caused by problems of translation between forms of life as trading zones; we note that where there is no problem of translation there is merely unproblematic 'trade'. Galison's supposed resolution to the general problem was to posit the existence of in-between languages – creoles and pidgins – which developed to enable 'trade' to happen between communities with radically different languages. His paradigm case was the invention of biochemistry which grew out of the invention of a new language which captured the appropriate parts of the language of chemistry and the language of biology. Galison's resolution seems to work well for this and similar cases but we argue in our earlier paper that this is just one way in which the problems of trading zones are resolved and there are many others. These many different ways are represented in Figure 1 in Collins et al 2007; there they are organised in a 2x2 table. This is reproduced as the shaded part of Figure 1 below.

As can be seen, one dimension of the 2x2 table is the final degree of melding, or homogeneity, of the communities (high in the case of biochemistry) while the other dimension is the degree of coercion used to bring about the cooperation (low in the case of biochemistry). As can be seen, biochemistry is one example in the top left-hand cell representing 'inter-language' trading zones. For the complete explication of this scheme readers should go back to the original paper.

## 2. Some additions to the original model

Here we make four additions to the scheme just described; these are shown in the right-hand unshaded column of Figure 1. The dashed lines and arrows indicate that the additions belong in the two right hand cells of the original table. We believe that with these additions the scheme better represents the full range of different ways in which cross-linguistic-community communication takes place in practice.

---

[1] The sources for forms of life are Winch, 1958, and Wittgenstein, 1953. Discussion of the 'fractal model' of forms of life in which they are embedded within one another and overlap can be found in Collins, 2011.

*Figure 1: Original model of trading zones (shaded) with additional categories (unshaded)*

In the top right hand, relatively voluntary, 'fractionated trading zone', cell (as well as the enforced cell below it), the parties to a cross-disciplinary interaction, rather than attempting a merger, continue to maintain the difference between their forms-of-life even while they cooperate; this is in contrast to the 'inter-language cell'. Joint work in the fractionated cell is achieved by the parties sharing only a fraction of their respective forms-of-life. The left-most method in this cell is sharing a 'boundary object', but here we extend the rightmost method, interactional expertise, when the fraction that is shared is the 'practice language' with no attempt being made to share practices.[2] In other words, parties try to learn the linguistic discourse of those with whom they wish to work without trying to engage in the same activities. The first extension to this cell is the 'ambassadorial model'.

---

[2] The term 'boundary object' is often used rather loosely and care should be taken to make sure that real explanatory work is being done when the term is invoked.

## 2.1 Ambassadorial model

In the original paper it is assumed that the working of the interactional expertise method depended on every member of each community learning the other party's language. But another important method is the ambassadorial model. Here, rather than all members trying to learn the new practice language, one or more individuals from the 'home group' are selected to spend enough time with the 'away group' to master the target interactional expertise. They can then *represent* the thinking of the other group within the home group. *Representing* is not the same as translating; translation is always incomplete if not impossible across deep cultural divides.

An example of the ambassadorial model at work can be seen in gravitational wave physics research.[3] In this research it is vital that potential detections are promulgated to the regular astronomical community so that they can point their telescopes in hope of seeing light or radio signals that correspond to a putative gravitational wave signal. Such a signal might be caused by the coalescence of two binary stars but probably not by the coalescence of two black holes – this is a matter of astrophysics. Members of the gravitational wave community are sent to spend time with the astronomical and astrophysical communities to learn their ways of thinking and these ambassadors can represent the astronomers and astrophysicists in the gravitational wave group as the protocols for joint observation are worked out – they can say such things as 'this is how the astronomers will think or react to that suggestion and this is what they would prefer'. Here is an example of such a phrase from an email circulated to the gravitational wave physics community on December 24th 2016:

> I concur with XXXX et al. that our EM partners would prefer we send out more triggers than less. ['EM' stands for 'electromagnetic' and refers to regular astronomers who mostly look for electromagnetic signals rather than gravitational wave signals.]

---

[3] For gravitational wave physics see, for example, Collins, 2004, 2017.

Ambassadors could also be sent the other way – from astronomy and from astrophysics to the gravitational wave community.

**2.2 Referred expertise model**

A second, closely related, addition to this cell is the referred expertise model. Here one-time astrophysicists (more likely than astronomers), will become members of the gravitational wave community, bringing their expertise with them. We call this an example of 'fractionated' cooperation because we are thinking of the recruited astrophysicists as *representing* their old community and fitting in via their newly learned *interactional expertise* in gravitational wave physics. This then is the astrophysicists acting like emissaries, rather than continuing to practice their old expertises within the new community. The difference is that the referred experts learn their trade in their home community before travelling whereas the case of the ambassadorial model, the ambassadors undertake an expedition to someone else's community to learn a new trade. It is worth noting, however, that in gravitational wave physics both the ambassadorial and the referred expertise models are temporary phenomena – they apply only to the pre-detection era when astrophysics and astronomy were clearly distinct from gravitational wave physics. In that era gravitational wave detection was *physics* – something that gave rise to much bad feeling with the first large interferometric detectors referred to themselves as 'LIGO' – standing for 'Laser Interferometer Gravitational-Wave

Observatory'; astronomers complained that it was not an observatory but a physics experiment. Now that gravitational waves have been detected, however, with many more observations expected shortly, LIGO *has* become an observatory and the distinction between the detection of gravitational waves, on the one hand, and astronomy and astrophysics, on the other, is disappearing. Gravitational wave physics is becoming part of astronomy and astrophysics and one can already see the first signs of the two communities merging and the cultures becoming unified (see Collins, 2017). It will soon cease to be correct to think of their being a trading zone, or even *trade* linking the enterprises as there will be no enterprises, only an enterprise.

**2.3 Deliverables model**

With 'deliverables' thought of as a means of communication, we move away from the fractionated trading zone cell to the, bottom right, 'enforced' cell. Specified deliverables follow the same cognitive model as the galley slaves, discussed in Collins et al 2007, who might not even know that they are propelling a ship so long as they pull on a pole in response to the slave master's punishments and rewards. The analogy is brutal but it is useful because it makes the cognitive model of deliverables clear; the only people doing any of the understanding required to meld the deliverables into the home activity are the home group. Of course, payment for services rendered is not as brutal as slavery but in both cases the provider of services – slave or consultant – need have no idea what they are doing so long as they deliver the specified object.

In the gravitational wave field the task required might be something like, 'install a seismometer of sensitivity such-and-such at such and such a location with readout that can be fed into a computer'. The seismometer installer need not know that the location is close to a scientific instrument; need not know that the instrument is an interferometer; need not know that the interferometer is part of a network of gravitational wave detectors; and need not know that the readings will be used to 'veto' stretches of potential signal that are contaminated by seismic noise.

It is probable that many difficulties emerge from confusing short-term referred expertise with deliverables. For example, we, the Cardiff expertise group, employed a software firm to build a program to run our Imitation Game experiments.[4] We assumed that the firm would make adjustments in response to their growing understanding of our needs – using their expertise in our context. We found, however, that they interpreted their job as doing only what we could formally specify in advance with any departures that came with growing understanding of the task in its context being rejected except on pain of extra charges. In research it is impossible to specify everything in advance so what we really needed was referred expertise, not specified deliverables, and we have consequently employed our own programmer to finish the job and that programmer has become part of our research team. There is a useful lesson here for the relationship between software houses and customers and all such contractual arrangements. Wherever the desired expertise is less than completely specifiable at the outset, it should be *absorbed* in one way or another rather than purchased.

**2.4 Multi-disciplinary model**

Multi-disciplinarity is an extension of the specified deliverable model. It differs from *inter*disciplinarity because there is no attempt at common understanding by either home group or foreign group – indeed, it is not clear if there is a 'home' group. Under *multi*disciplinarity many deliverers, or groups of deliverers, are brought together to contribute their skills to some project without understanding the overall goal or their contribution to it.

---

[4] For Imitation Games see, for example, Collins and Evans, 2014

There is no real trading zone. Mass slave labour is the cognitive model. Once more, someone has to understand how all the deliverables fit together if such projects are to work – which they often do not. Since none of the cooperating parties understand the other parties at any deep level, the 'slave-master' – the manager who is holding the whole project together – will have to be a person of remarkable abilities with at least interactional expertise in every separate discipline represented. Tragically, multidisciplinary projects are usually funded and organised on the assumption that the contributions of all the separate disciplines will slot together automatically like the pieces of a jigsaw puzzle. This is hardly ever going to be the case.

## 3. Conclusion

In this paper we have elaborated the model of trading zones first presented in Collins et al 2007. We believe this expanded version of the model includes some very important but previously overlooked ways for separate language communities to communicate. We have found examples that fit three out of the four additional cases and have offered some insights into how to avoid at least one kind of failure, but we are pessimistic about the likelihood of success of the fourth model.